\documentclass[secnumarabic,12pt,tightenlines,eqsecnum,floats,aps,amsmath,amssymb,nofootinbib,superscriptaddress,showpacs,prd]{revtex4-1}

\usepackage{amsmath,amsthm,amssymb,amsfonts}
\usepackage{graphicx}
\usepackage{supertabular}
\usepackage{mathcomp}
\usepackage{float}
\usepackage{units}
\usepackage{hyperref}
\usepackage{bm}
\usepackage{mathtools}

\numberwithin{equation}{section}

\begin{document}

\title{Gauge invariance in simple mechanical systems}

\author{J. Fernando \surname{Barbero G.}}
\email[]{fbarbero@iem.cfmac.csic.es}
\affiliation{Instituto de Estructura de la Materia, CSIC, Serrano 123, 28006 Madrid, Spain}
\affiliation{Grupo de Teor\'{\i}as de Campos y F\'{\i}sica Estad\'{\i}stica, Instituto Universitario Gregorio Mill\'an
Barbany, Universidad Carlos III de Madrid, Unidad Asociada al IEM-CSIC.}

\author{Jorge \surname{Prieto}}
\email[]{jorgeprietoarranz@gmail.com}
\affiliation{Instituto Gregorio Mill\'an, Grupo de Modelizaci\'on y Simulaci\'on Num\'erica, Universidad Carlos III de Madrid, Avda. de la Universidad 30, 28911 Legan\'es, Spain}

\author{Eduardo J. \surname{S. Villase\~nor}}
\email[]{ejsanche@math.uc3m.es}
\affiliation{Instituto Gregorio Mill\'an, Grupo de Modelizaci\'on y Simulaci\'on Num\'erica,
Universidad Carlos III de Madrid, Avda. de la Universidad 30, 28911 Legan\'es, Spain}
\affiliation{Grupo de Teor\'{\i}as de Campos y F\'{\i}sica Estad\'{\i}stica, Instituto Universitario Gregorio Mill\'an Barbany, Universidad Carlos III de Madrid, Unidad Asociada al IEM-CSIC.}

\date{January 9, 2015}

\begin{abstract}
This article discusses and explains the Hamiltonian formulation for a class of simple gauge invariant mechanical systems consisting of point masses and idealized rods. The study of these models may be helpful to advanced undergraduate or graduate students in theoretical physics to understand, in a familiar context, some concepts relevant to the study of classical and quantum field theories. We use a geometric approach to derive the Hamiltonian formulation for the model considered in the paper: four equal masses connected by six ideal rods. We obtain and discuss the meaning of several important elements, in particular, the constraints and the Hamiltonian vector fields that define the dynamics of the system, the constraint manifold, gauge symmetries, gauge orbits, gauge fixing, and the reduced phase space.
\end{abstract}

\maketitle

\section{Introduction}

Gauge field theories play a central role in the description of the fundamental interactions of physics. A popular way to present the concept of gauge invariance is based on the idea of turning \textit{global} symmetries into \textit{local} ones, involving arbitrary functions, through the introduction of the so called gauge fields. In many contexts gauge theories are defined, more or less explicitly, precisely as those obtained by following this procedure. Their physical usefulness hinges upon the possibility of finding \textit{observables} that are insensitive to the presence of these arbitrary elements; i.e. the identification of suitable gauge invariant functions of the dynamical variables.

An indirect consequence of the introduction of local symmetries is the fact that the field equations become singular. This singularity manifests itself as the impossibility to solve for some of the second order time derivatives of the fields in terms of the other objects present in the equations of motion. An associated effect is the possible appearance of arbitrary functions in their solutions (notice, though, the existence of singular systems such as the Proca field for which no arbitrariness shows up). From this perspective gauge theories are a particular instance of the more general models described by singular Lagrangians (i.e. those leading to singular Euler-Lagrange equations).

The traditional way to deal with singular Lagrangians and the canonical quantization of the physical models defined by them relies on the ideas developed by Dirac \cite{Dirac}. A key feature of the algorithm introduced by him to get the Hamiltonian formulation for these systems is the appearance of constraints, i.e. conditions that the canonical variables must satisfy at all times during the evolution. The quantized version of the phase space functions that represent these constraints is a key element in Dirac's approach.

A common and widespread misconception is to think that gauge theories must necessarily involve fields, reparametrizations and changes of coordinates or reference frames. The purpose of this paper is to show that the dynamics of simple mechanical systems, consisting of a finite number of point particles connected by ideal rods, can display gauge behavior. In order to make our presentation as pedagogical as possible we will focus on a specific example, consisting of four equal masses connected by six rods, and compare it with the quintessential gauge theory: electromagnetism (EM). We will analyze in detail the Lagrangian and Hamiltonian formulations for the particular model considered here. As we will show our example mimics some of the crucial features of electromagnetism and is richer in some sense, in particular regarding its Hamiltonian formulation.

The best way to understand the essence of Dirac's construction is in geometric terms. A very clear perspective on this issue was provided by Gotay, Nester and Hinds  (GNH)  \cite{GNH,Gotay} so we will use their method. Instead of giving an abstract description of the GNH algorithm we will introduce it as we perform the actual computations for our model. In our opinion the present paper will serve the dual purpose of clarifying some of the concepts behind gauge systems in a very simple setting (constraints, gauge orbits, gauge symmetry, gauge fixing, reduced phase space...) and also provide a pedagogical introduction to the Hamiltonian description of singular systems.

The structure of the paper is the following. After this introduction we will start to study in section \ref{fourparticle-sect} a particular, but representative, model that displays the gauge behavior that we want to discuss: four particles connected by six rods. Section \ref{fourparticle-2-sect} will be devoted to obtaining the Hamiltonian formulation for this system in a neat way by using geometric methods inspired in the GNH algorithm. The paper ends in section \ref{conclusions-sect} with a short discussion.

A comment is in order here; in order to make the paper accessible to advanced undergraduate students we are not assuming any prior knowledge of differential geometry on the part of the reader (only standard multivariate calculus), however, we will mention by name some of the relevant geometric objects to justify their use and show their logical connection with the concepts discussed in the paper. We will gloss over several technical points that can be skipped in a first approach to this subject. Readers interested in the geometrization of classical dynamics are referred to the comprehensive treaty by Abraham and Marsden \cite{Abraham-Marsden}.

\section{The four particle model}\label{fourparticle-sect}

Singular finite dimensional dynamical systems have been considered in some detail in the literature (models displaying different pathological behaviors can be found, for instance, in the book by Henneaux and Teitelboim \cite{HT}). The main drawback of the usual examples is their rather artificial character, i.e. they do not describe systems with a simple physical interpretation. One of the goals of this paper is to provide one such example. We base our approach on the possibility of considering some constraint forces as dynamical variables on a par with the standard (generalized) coordinates. The other central idea is to reproduce, to some extent, the behaviour of hyperstatic systems\footnote{These are structures for which the equations of statics do not suffice to determine all the internal forces.} in a \textit{dynamical situation}.

The simplest model that we could discuss would be a system of two point masses connected by two ideal rods (massless and completely rigid). It is obvious that the force exerted by each rod is undetermined: only their sum can have a physical meaning. Similar arrangements with an arbitrary number $N$ of rods or other collinear models display the same kind of behavior. We will discard them here for two main reasons: they are somehow trivial on one hand (i.e. the $N$-rod model with two masses) or non-generic in a concrete sense (collinear systems are \textit{infinitesimally flexible} in the parlance of reference \cite{Roth}). Furthermore, the natural representation of our model, inspired in graph theory, can be generalized to the study of the dynamics of more interesting and non trivial examples but is not suitable for systems with collinear rods.

\begin{figure}[h!]
\centering
\includegraphics[width=9cm]{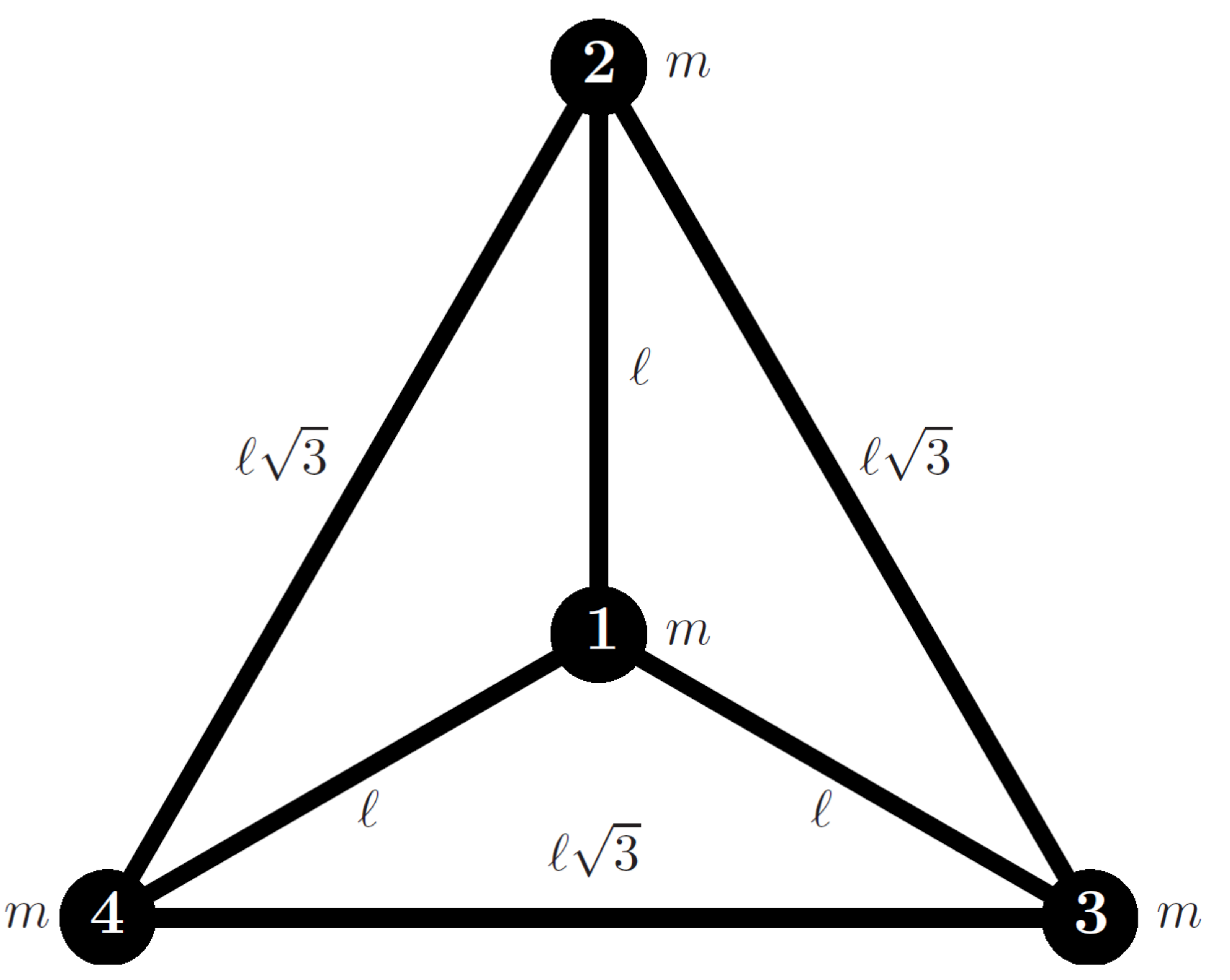}
\caption{The four masses of magnitude $m$ are connected with six rigid rods to form an equilateral triangle of side $\ell\sqrt{3}$. The central mass is placed at the barycenter of the triangle.}
\label{4masses6rods}
\end{figure}

Our model consists of four equal masses in a triangular arrangement (the fourth mass is placed at the barycenter) connected by six rods as shown in Fig. \ref{4masses6rods}. The system is constrained to move in the plane. Notice that the structure is rigid and remains so even after removing one of the rods. We take the following Lagrangian
\begin{eqnarray}
L(\mathbf{q}_i,\!q_{\{i,j\}},\!\mathbf{v}_i,\!v_{\{i,j\}}\!)&\!=\!&\frac{1}{2}m\sum_{i\in \mathcal{V}}\|\mathbf{v}_i\|^2
-\frac{1}{2}\sum_{\{i,j\}\in \mathcal{E}}\!\!q_{\{i,j\}}\big(\|\mathbf{q}_i-\mathbf{q}_j\|^2-\ell^2_{\{i,j\}}\big)\,,\label{Lagrangian}
\end{eqnarray}
where our notation makes use of the graph labels associated with our system according to the numbering shown in Fig. \ref{4masses6rods}.
Here the sets of vertices and edges are respectively given by $\mathcal{V}=\{1,\ldots,4\}$ and $\mathcal{E}=\big\{\{1,2\},\{1,3\},\{1,4\},\{2,3\},\{2,4\},\{3,4\}\big\}$. We denote the length of the $\{i,j\}$ edge as $\ell_{\{i,j\}}$ (see Fig. \ref{4masses6rods} for their values). The positions and velocities of the particles are represented by $\mathbf{q}_i$ and $\mathbf{v}_i$ respectively. Each of the configuration variables $q_{\{i,j\}}$ is a Lagrange multiplier enforcing the condition that the length of the edge $\{i,j\}$ is $\ell_{\{i,j\}}$. These conditions are holonomic constraints (i.e. velocity independent) and, hence, can be introduced in the Lagrangian in this simple way without modifying the Newtonian equations of motion. The velocities associated with the $q_{\{i,j\}}$ are denoted as $v_{\{i,j\}}$.

We pause for a moment to compare $L$ with other well known Lagrangians, in particular the one describing the free electromagnetic field given by
\begin{eqnarray}
L_{EM}(\mathbf{A},A_0,\mathbf{V},V_0)&\!=\!&\frac{1}{2}\int  \Big((\mathbf{V}+\boldsymbol{\nabla} A_0)\cdot(\mathbf{V}+\boldsymbol{\nabla} A_0)
-(\boldsymbol{\nabla}\times\mathbf{A})\cdot(\boldsymbol{\nabla}\times\mathbf{A})\Big)\mathrm{d} \mathbf{x}\,.\label{LagrangianEM}
\end{eqnarray}
Here $\mathbf{A}$ and $A_0$ are functions\footnote{Their dependence on the spatial coordinates can be roughly understood as the presence of a continuous index in analogy with the index $i\in \mathcal{V}$ that labels the particles in our model.}  on $\mathbb{R}^3$ that denote the vector and scalar potentials respectively with the corresponding velocities being $\mathbf{V}$ and $V_0$. As we can see $L$ and $L_{EM}$ share some features, for instance, the velocities associated with some of the dynamical variables ($q_{\{i,j\}}$ and $A_0$) do not appear. Although the introduction of the terms involving $q_{\{i,j\}}$ in $L$ may seem as an artificial complication they are, in fact, similar to the ones involving $A_0$ in $L_{EM}$. We exploit this analogy in the present paper.

The Euler-Lagrange equations derived from the Lagrangian $L$ give
\begin{equation}
m\ddot{\mathbf{q}}_i=-\sum_{j\sim i}q_{\{i,j\}}(\mathbf{q}_i-\mathbf{q}_j)\,,\quad i\in \mathcal{V}\,,
\label{qequations}
\end{equation}

\vspace*{-4mm}

\begin{equation}
\|\mathbf{q}_i-\mathbf{q}_j\|^2=\ell^2_{\{i,j\}}\,,\hspace*{1.6cm}\{i,j\}\in \mathcal{E}\,,\label{lagequations}
\end{equation}
where the $j\sim i$ notation means that the sum extends over all vertices $j$ connected with the fixed vertex $i$.

Notice that each term $-q_{\{i,j\}}(\mathbf{q}_i-\mathbf{q}_j)$ in the r.h.s of Eq. \eqref{qequations} can be interpreted as the force exerted on the particle $i$ by the rod connecting it with particle $j$ because the sum of these terms \textit{is} the force acting on the $i$th particle. As there are no terms involving $\ddot{q}_{\{i,j\}}$ the system is singular. It is easy to show the existence of families of solutions to these equations that depend on a free function but describe the same physics. Consider for instance
\begin{subequations}
\label{solution_qi}
\begin{align}
\mathbf{q}_1(t) &\;=\;(0,0)\,,\label{q1}\\
\mathbf{q}_2(t) &\;=\;(-\ell\sin\omega t,\ell\cos\omega t)\,,\label{q2}\\
\mathbf{q}_3(t) &\;=\;(-\ell\sin(\omega t-2\pi/3),\ell\cos(\omega t-2\pi/3))\,,\label{q3}\\
\mathbf{q}_4(t) &\;=\;(-\ell\sin(\omega t+2\pi/3),\ell\cos(\omega t+2\pi/3))\,,\label{q4}
\end{align}
\end{subequations}
and
\begin{subequations}
\label{qij}
\begin{align}
q_{\{1,2\}}(t)& =  q_{\{1,3\}}(t)=q_{\{1,4\}}(t) =f(t)\,,\label{q12}\\
q_{\{2,3\}}(t)& = q_{\{2,4\}}(t)=q_{\{3,4\}}(t)=\frac{1}{3}(m\omega^2-f(t)) \,,\label{q23}
\end{align}
\end{subequations}
where $f(t)$ is an \textit{arbitrary} function of time and $\omega$ a real parameter. It is straightforward to check that these functions satisfy Eqs. \eqref{qequations} and \eqref{lagequations}. Notice that the motion of each particle is perfectly determined, which implies that the force acting on each of them also is. However, the force exerted by each rod \textit{is not completely} determined. This is exactly what happens in gauge theories: some magnitudes are arbitrary to a certain degree but there are ``physical functions'' of them (observables) that are not arbitrary.

The constraints \eqref{lagequations} imply that we cannot freely choose initial positions for all the particles in the system. They also imply that the velocities cannot be arbitrary either.\footnote{The velocity field for a rigid solid has a very specific form that determines the actual freedom to choose the initial velocities of the particles.} A set of conditions that the velocities must satisfy can be obtained by differentiating Eq. \eqref{lagequations}:
\begin{equation}
(\mathbf{q}_i-\mathbf{q}_j)\cdot(\dot{\mathbf{q}}_i-\dot{\mathbf{q}}_j)=0\,,\quad \{i,j\}\in \mathcal{E}\,.
\label{velconstraints}
\end{equation}
At this point a simple procedure to obtain all the possible conditions on the configuration variables and velocities suggests itself: keep on differentiating and using, when possible, the equations of motion and the constraints already obtained to determine if new independent conditions appear. Although this method can actually be implemented in the present case\footnote{The Lagrangian symplectic approach \cite{Gotay} would provide the rigorous geometric implementation of this procedure.} there are two reasons not to do so. The first is that it is somehow difficult to find out when the procedure stops. The second is that we are interested in the Hamiltonian formulation (necessary, for example, to attempt the canonical quantization of our system \textit{\`a la Dirac}).

\section{The four particle model: the Hamiltonian picture}\label{fourparticle-2-sect}

We obtain now the Hamiltonian formulation for our system by using a method inspired in the GNH algorithm \cite{GNH}. The first step to get the Hamiltonian formulation for a mechanical model from its Lagrangian is to define the canonical momenta and write the generalized velocities in terms of them. Singular systems are identified, in practice, as those for which this is not possible. One might hastily conclude that the Hamiltonian formulation does not exist in this case; however, a quick look at the equations of motion suggests a possible way out: the fact that the positions of the particles and their velocities are subject to constraints such as \eqref{lagequations} or \eqref{velconstraints} could mean that the proper space to define the Hamiltonian dynamics is not the ``full phase space'' $\Gamma$ consisting of all the generalized positions $\mathbf{q}_i$, $q_{\{i,j\}}$ and momenta $\mathbf{p}_i$, $p_{\{i,j\}}$ but rather an appropriate subset of it. The definition of the canonical momenta $p=\partial L/\partial v$ from the Lagrangian $L$ given by Eq. \eqref{Lagrangian}, can be interpreted as a map (known in the technical literature \cite{Abraham-Marsden} as the \textit{fiber derivative}) $FL:\mathbb{R}^{28}\rightarrow \Gamma $ defined by
\begin{eqnarray}
(\mathbf{q}_i,q_{\{i,j\}},\mathbf{v}_i,v_{\{i,j\}})\mapsto (\mathbf{q}_i,q_{\{i,j\}},\partial L/ \partial \mathbf{v}_i,\partial L/\partial v_{\{i,j\}})=(\mathbf{q}_i,q_{\{i,j\}},m\mathbf{v}_i,0)\,.\label{Fiberderivative}
\end{eqnarray}
As the momenta $p_{\{i,j\}}$ are zero, the image under $FL$ of a curve $(\mathbf{q}_k(t),q_{\{i,j\}}(t),\dot{\mathbf{q}}_k(t),\dot{q}_{\{i,j\}}(t))$ must be contained in the so called \textit{primary constraint submanifold} of $\Gamma$ given by
\begin{equation}
\mathcal{M}_0:=\{(\mathbf{q}_i,q_{\{i,j\}},\mathbf{p}_i,p_{\{i,j\}})\in\mathbb{R}^{28}:\mathcal{C}^{(0)}_{\{i,j\}}:=p_{\{i,j\}}=0\,,\{i,j\}\in \mathcal{E}\}\,.
\label{M0}
\end{equation}
It is, hence, natural to look for a Hamiltonian description defined only on $\mathcal{M}_0$ or an appropriate subset of it. $\mathcal{M}_0$ can be viewed as $\mathbb{R}^{22}$ with coordinates $(\mathbf{q}_i,q_{\{i,j\}},\mathbf{p}_i)$.

The evolution of the system in Hamiltonian form is given by the integral curves parameterized by time $(\mathbf{q}_i(t),q_{\{i,j\}}(t),\mathbf{p}_i(t),p_{\{i,j\}}(t))$  of a vector field $\mathbf{X}=(\mathbf{X}_{\mathbf{q}_i},X_{q_{\{i,j\}}},\mathbf{X}_{\mathbf{p}_i},X_{p_{\{i,j\}}})$. These are given by the first order differential equations
\begin{equation}
\dot{\mathbf{q}}_i=\mathbf{X}_{\mathbf{q}_i}\,,\dot{q}_{\{i,j\}}=X_{q_{\{i,j\}}}\,,\dot{\mathbf{p}}_i=\mathbf{X}_{\mathbf{q}_i}\,,\dot{p}_{\{i,j\}}=X_{p_{\{i,j\}}}\,.
\label{integralcurves}
\end{equation}
The vector field $\mathbf{X}$ is obtained by using a construction that involves the Hamiltonian $H$ and an important geometric object: the \textit{symplectic form} $\boldsymbol{\Omega}$.

The usual prescription to obtain a Hamiltonian  (based on performing a Legendre transform) cannot be applied to this example but a simple extension of it can be used on $\mathcal{M}_0$. In the present case this amounts to ignoring the $p_{\{i,j\}}\dot{q}_{\{i,j\}}$ terms to get
\begin{eqnarray}
H(\mathbf{q}_i,q_{\{i,j\}},\mathbf{p}_i)&=&\frac{1}{2m}\sum_{i\in \mathcal{V}}\|\mathbf{p}_i\|^2
+\frac{1}{2}\sum_{\{i,j\}\in \mathcal{E}}q_{\{i,j\}}\big(\|\mathbf{q}_i-\mathbf{q}_j\|^2-\ell^2_{\{i,j\}}\big)\,.\label{Hamiltonian}
\end{eqnarray}
Although this function is defined in principle only on $\mathcal{M}_0$ it can be trivially extended to $\Gamma$ as $H(\mathbf{q}_i,q_{\{i,j\}},\mathbf{p}_i,p_{\{i,j\}})=H(\mathbf{q}_i,q_{\{i,j\}},\mathbf{p}_i)$.

The inhomogeneous linear equation that would determine $\mathbf{X}$ if the system were non-singular is\footnote{This form is enough for our needs in this paper. A more rigorous way to write this equation requires the use of differential forms and related concepts.}
\begin{equation}
\boldsymbol{\Omega} \mathbf{X}-\boldsymbol{\nabla} H=0\,,\label{definition_X}
\end{equation}
where the $28$ entries of $\boldsymbol{\Omega} \mathbf{X}$ (defined by the linear action of $\boldsymbol{\Omega}$ on $\mathbf{X}$) are real functions on the phase space and $\boldsymbol{\nabla} H$ denotes the $28$-dimensional gradient of $H$. The geometric structure of $\Gamma$ is such that a specific symplectic form $\boldsymbol{\Omega}$ with the required properties (non-degeneracy among them) can always be built ---hence the word \textit{canonical}. A famous theorem by Darboux \cite{Abraham-Marsden} proves that it is always possible to find a coordinate system covering a large enough part of $\Gamma$ (an open set) where $\boldsymbol{\Omega}$ can be written in matrix form as
\begin{equation}
\begin{pmatrix}
	 \ \ 0_{14\times14} & \ -I_{14\times14} \\
	I_{14\times14} & \ 0_{14\times14} \\
\end{pmatrix}.
\end{equation}
This is actually the reason why elementary treatments avoid discussing the determination of the vector field $\mathbf{X}$ from the Hamiltonian through the solution of \eqref{definition_X}: when $\boldsymbol{\Omega}$ takes the previous form the equations for the integral curves of $\mathbf{X}$ are the textbook Hamilton equations. The Hamiltonian treatment of singular systems requires in an unavoidable way the explicit consideration of the symplectic structure. This is why we mention it here.

Given that the dynamics in our example must be confined to $\mathcal{M}_0$ ($p_{\{i,j\}}=0$), we can work as if this was the full phase space, in particular, try to find Hamiltonian vector fields $\mathbf{X}=(\boldsymbol{X},0)=(\mathbf{X}_{\mathbf{q}_i},X_{q_{\{i,j\}}},\mathbf{X}_{\mathbf{p}_i},0)$, i.e. tangent to $\mathcal{M}_0$ and defined only there. Acting on these vectors Eq. \eqref{definition_X} becomes
\begin{equation}
(\boldsymbol{\omega} \boldsymbol{X}-\boldsymbol{\nabla} H)|_{\mathcal{M}_0}=0\,,\label{definition_X_restricted}
\end{equation}
where $\boldsymbol{\omega}$ is the $22\times22$ degenerate matrix
\begin{equation}
\boldsymbol{\omega}=
\begin{pmatrix}
	 0_{8\times 8} & \ \ 0_{8\times 6} & -I_{8\times 8} \\
	 0_{6\times 8} & \ \ 0_{6\times 6}  & \  0_{6\times 8}\\
     I_{8\times 8} & \ \ 0_{8\times 6}  & \ 0_{8\times 8}\\
\end{pmatrix}\,.
\label{pulbbackomega}
\end{equation}
The action of $\boldsymbol{\omega}$ on $\boldsymbol{X}$ that we need in order to solve Eq. \eqref{definition_X_restricted} is
\begin{equation}
\boldsymbol{\omega} \boldsymbol{X}=(-\mathbf{X}_{\mathbf{p}_i},\boldsymbol{0}_6,\mathbf{X}_{\mathbf{q}_i})\,.
\label{OmegaX}
\end{equation}
The gradient on $\mathcal{M}_0$ is $\boldsymbol{\nabla}:=(\partial_{\mathbf{q}_i},\partial_{q_{\{i,j\}}},\partial_{\mathbf{p}_i})$, hence,
\begin{equation}
\boldsymbol{\nabla} H\!\!=\!\big(\!\sum_{j\sim i}q_{\{i,j\}}(\mathbf{q}_i-\mathbf{q}_j),\frac{1}{2}(\|\mathbf{q}_i-\mathbf{q}_j\|^2\!\!-\ell^2_{\{i,j\}}),\frac{\mathbf{p}_i}{m}\big)\,.
\label{gradientH}
\end{equation}
The equations (\ref{definition_X_restricted}) constitute a linear, inhomogeneous system so, generically, some condition must be satisfied by the inhomogeneous term for the system to be solvable. It may also happen that only part of the unknowns (the components of $\boldsymbol{X}$ in this example) are fixed after solving it. In the present case we easily obtain
\begin{eqnarray}
&&\mathbf{X}_{\mathbf{q}_i}=\frac{\mathbf{p}_i}{m}\,,\hspace*{5cm} i\in \mathcal{V}\label{Xq}\\
&&\mathbf{X}_{\mathbf{p}_i}=-\sum_{j\sim i}q_{\{i,j\}}(\mathbf{q}_i-\mathbf{q}_j)\,,\hspace*{1.95cm} i\in \mathcal{V}\label{Xp}\\
&&0=\|\mathbf{q}_i-\mathbf{q}_j\|^2-\ell^2_{\{i,j\}}=:\mathcal{C}^{(1)}_{\{i,j\}}\,,\quad \{i,j\}\in \mathcal{E}\,.\label{SecondaryConstrain1}
\end{eqnarray}
As we can see Eq. (\ref{definition_X_restricted}) cannot be solved on the whole of $\mathcal{M}_0$ but only in the part of it where condition (\ref{SecondaryConstrain1}) is satisfied. Let us call this subset $\mathcal{M}_1$, i.e.
\begin{equation}
\mathcal{M}_1:=\{(\mathbf{q}_i,q_{\{i,j\}},\mathbf{p}_i)\in\mathcal{M}_0:\mathcal{C}^{(1)}_{\{i,j\}}=0\,,\{i,j\}\in \mathcal{E}\}\,.
\label{M1}
\end{equation}
The components of $\boldsymbol{X}$ must have the form given by Eqs. \eqref{Xq} and \eqref{Xp}. Notice that the components $X_{q_{\{i,j\}}}$ remain arbitrary at this stage. Now, if the vector field $\boldsymbol{X}$, with the form just obtained, were \textit{tangent} to $\mathcal{M}_1$ we would have succeeded in finding an appropriate submanifold of the phase space where we can define the Hamiltonian dynamics of our singular system. This can be shown by checking if the directional derivative $\boldsymbol{\nabla}_{\boldsymbol{X}} \mathcal{C}^{(1)}_{\{i,j\}}$ of $\mathcal{C}^{(1)}_{\{i,j\}}$ along $\boldsymbol{X}$ \emph{vanishes}. In this case we have
\begin{equation}
\boldsymbol{\nabla}_{\boldsymbol{X}} \mathcal{C}^{(1)}_{\{i,j\}}=\frac{2}{m}(\mathbf{q}_i-\mathbf{q}_j)\cdot(\mathbf{p}_i-\mathbf{p}_j)=:\mathcal{C}^{(2)}_{\{i,j\}}\,,
\label{SecondaryConstrain2}
\end{equation}
which implies that $\boldsymbol{X}$ is only tangent to $\mathcal{M}_1$ at the points satisfying $\mathcal{C}^{(2)}_{\{i,j\}}=0$. These define the new submanifold
\begin{equation}
\mathcal{M}_2:=\{(\mathbf{q}_i,q_{\{i,j\}},\mathbf{p}_i)\in\mathcal{M}_1:\mathcal{C}^{(2)}_{\{i,j\}}=0\,,\{i,j\}\in \mathcal{E}\}\,.
\label{M2}
\end{equation}
The conditions $\mathcal{C}^{(2)}_{\{i,j\}}=0$ are necessarily independent of (\ref{SecondaryConstrain1}) as they involve the momenta $\mathbf{p}_i$. It is interesting at this point to pause for a moment to understand their meaning. First of all, as the system is contained in the plane, it is obvious that
\begin{equation}
(\mathbf{q}_i-\mathbf{q}_j)\cdot(\mathbf{p}_i-\mathbf{p}_j)=0\Leftrightarrow \mathbf{p}_i-\mathbf{p}_j=\omega_{\{i,j\}} R(\mathbf{q}_i-\mathbf{q}_j)\,,\quad \{i,j\}\in \mathcal{E}\,,
\label{prigidbody}
\end{equation}
where $R$ is a counterclockwise rotation of $\pi/2$ and $\omega_{\{i,j\}}$ are real coefficients. If we select three of the particles in the system (say, 1, 2 and 3) and add the expressions given in Eq. \eqref{prigidbody} for $\{i,j\}=\{1,2\},\{2,3\}$ and $\{1,3\}$ we immediately see that $\omega_{\{1,2\}}=\omega_{\{2,3\}}=\omega_{\{1,3\}}=:m\omega$ with $\omega\in\mathbb{R}$. Considering the remaining triangles in the graph associated with the system we get $\omega_{\{i,j\}}=m\omega$ for every $\{i,j\}\in \mathcal{E}$, that is, the constraints $\mathcal{C}^{(2)}_{\{i,j\}}=0$ are equivalent to the existence of a real parameter $\omega$ such that
\begin{equation}
\mathbf{p}_i-\mathbf{p}_j=m\omega R(\mathbf{q}_i-\mathbf{q}_j)\,,\quad \{i,j\}\in \mathcal{E}\,.
\label{sol1}
\end{equation}
As in the present case the velocities are just the momenta divided by $m$ these last conditions are equivalent to saying that the velocities correspond to those of the particles of a rigid body (with an angular velocity given by $\omega$).

The way to proceed is obvious now, compute
\begin{equation}
\hspace*{-2mm}\boldsymbol{\nabla}_{\boldsymbol{X}} \mathcal{C}^{(2)}_{\{i,j\}}\!\!=\!\frac{1}{m}\|\mathbf{p}_i\!-\!\mathbf{p}_j\|^2\!-\!\!\sum_{k\sim i}\!q_{\{i,k\}}(\mathbf{q}_i\!-\!\mathbf{q}_j)\!\cdot\!(\mathbf{q}_i\!-\!\mathbf{q}_k)\!+\!\!\sum_{k\sim j}\!q_{\{j,k\}}(\mathbf{q}_i\!-\!\mathbf{q}_j)\!\cdot\!(\mathbf{q}_j\!-\!\mathbf{q}_k)\!=:\!\mathcal{C}^{(3)}_{\{i,j\}}\,,
\label{SecondaryConstrain3}
\end{equation}
and check if the conditions $\mathcal{C}^{(3)}_{\{i,j\}}=0$ for $\{i,j\}\in \mathcal{E}$ provide additional constraints. On the submanifold $\mathcal{M}_2$ these conditions can be written in the form
\begin{equation}
2m\omega^2\frac{\ell_{\{i,j\}}^2}{\ell^2}-\sum_{\{k,l\}\in \mathcal{E}}M_{\{i,j\}}^{\ \ \ \, \{k,l\}}q_{\{k,l\}}=0\,,\quad \{i,j\}\in \mathcal{E}\,
\label{Constraint3MatrixForm}
\end{equation}
where, by using the geometry of the system (see Fig. \ref{4masses6rods}), the matrix $\mathbf{M}=(M_{\{i,j\}}^{\ \ \ \, \{k,l\}})$ can be seen to be
\begin{equation}
\mathbf{M}=
\begin{pmatrix*}[r]
4&-1&-1&3&3&0\\
-1&4&-1&3&0&3\\
-1&-1&4&0&3&3\\
3&3&0&12&3&3\\
3&0&3&3&12&3\\
0&3&3&3&3&12
\end{pmatrix*}
\,.
\label{MatrixM}
\end{equation}
The entries of  $\mathbf{M}$ are labelled in the order $\{1,2\}$, $\{1,3\}$, $\{1,4\}$, $\{2,3\}$, $\{2,4\}$ and $\{3,4\}$. The rank of $\mathbf{M}$ is 5 and its kernel is spanned by the vector $\mathbf{u}^{\mathrm{T}}=(-3,-3,-3,1,1,1)$. It is straightforward to see that the conditions given by \eqref{Constraint3MatrixForm} are a \textit{compatible} system of equations for the $q_{\{i,j\}}$ where we can solve for any five of them in terms of the remaining one. The solutions of \eqref{Constraint3MatrixForm} can be parameterized, for example, in the form
\begin{equation}
q_{\{1,2\}}=q_{\{1,3\}}=q_{\{1,4\}}=\lambda,\quad q_{\{2,3\}}=q_{\{2,4\}}=q_{\{3,4\}}= -\frac{\lambda}{3} + \frac{ m \omega^2}{3}\,,\quad\lambda\in\mathbb{R}\,.\label{qijlambda}
\end{equation}
At this point we have the submanifold
\begin{equation}
\mathcal{M}_3:=\{(\mathbf{q}_i,q_{\{i,j\}},\mathbf{p}_i)\in\mathcal{M}_2:\mathcal{C}^{(3)}_{\{i,j\}}=0\,,\{i,j\}\in \mathcal{E}\}\,.
\label{M3}
\end{equation}
We need to check again if the vector field $\boldsymbol{X}$ is tangent to $\mathcal{M}_3$ by requiring $\boldsymbol{\nabla}_{\boldsymbol{X}} \mathcal{C}^{(3)}_{\{i,j\}}=0$, i.e.
\begin{eqnarray}
0&=&m \sum_{k\sim i}X_{q_{\{i,k\}}}(\mathbf{q}_i-\mathbf{q}_j)\!\cdot\!(\mathbf{q}_i-\mathbf{q}_k)
-m\sum_{k\sim j}X_{q_{\{j,k\}}}(\mathbf{q}_i-\mathbf{q}_j)\!\cdot\!(\mathbf{q}_j-\mathbf{q}_k)\nonumber\\
&&+\,3\, \sum_{k\sim i}q_{\{i,k\}}(\mathbf{p}_i-\mathbf{p}_j)\!\cdot\!(\mathbf{q}_i-\mathbf{q}_k)-\,3\, \sum_{k\sim j}q_{\{j,k\}}(\mathbf{p}_i-\mathbf{p}_j)\!\cdot\!(\mathbf{q}_j-\mathbf{q}_k)\label{lastsystem}\\
&&+\ \ \,\sum_{k\sim i}q_{\{i,k\}}(\mathbf{q}_i-\mathbf{q}_j)\!\cdot\!(\mathbf{p}_i-\mathbf{p}_k) -\ \ \,\sum_{k\sim j}q_{\{j,k\}}(\mathbf{q}_i-\mathbf{q}_j)\!\cdot\!(\mathbf{p}_j-\mathbf{p}_k)\,.\nonumber
\end{eqnarray}
By using again the geometry of the system and \eqref{sol1} we can write \eqref{lastsystem} in the form
\begin{equation}
\sum_{\{k,l\}\in \mathcal{E}}\!\!M_{\{i,j\}}^{\ \ \ \, \{k,l\}}X_{q_{\{k,l\}}}+2\sqrt{3}\,\omega\sum_{\{k,l\}\in \mathcal{E}}\!\!N_{\{i,j\}}^{\ \ \ \, \{k,l\}}\,\,q_{\{k,l\}}=0\,,\quad \{i,j\}\in E\,
\label{lastsystem2}
\end{equation}
where the matrix $\mathbf{N}=(N_{\{i,j\}}^{\ \ \ \, \{k,l\}})$ is
\begin{equation}
\mathbf{N}=
\begin{pmatrix*}[r]
0&-1&1&1&-1&0\\
1&0&-1&-1&0&1\\
-1&1&0&0&1&-1\\
-1&1&0&0&-3&3\\
1&0&-1&3&0&-3\\
0&-1&1&-3&3&0
\end{pmatrix*}
\,.
\label{MatrixN}
\end{equation}
This is a linear inhomogeneous system of equations for the $X_{q_{\{i,j\}}}$ components of the Hamiltonian vector field $\boldsymbol{X}$. No new constraints appear now as compatibility conditions because $\mathbf{u}^{\mathrm{T}}\mathbf{N}=0$. We can then solve for the \textit{functions} $X_{q_{\{i,j\}}}$ to finally get
\begin{equation}
X_{q_{\{1,2\}}}=X_{q_{\{1,3\}}}=X_{q_{\{1,4\}}}=\Xi,\quad X_{q_{\{2,3\}}}=X_{q_{\{2,4\}}}=X_{q_{\{3,4\}}}= -\frac{\Xi}{3}\,,\label{Xij}
\end{equation}
where we have used Eq. \eqref{qijlambda} and $\Xi$ is an arbitrary \textit{real function} on $\mathcal{M}_3$. The algorithm stops here because there are no more conditions on the canonical variables and we have been able to solve for the components of the vector field $\boldsymbol{X}$ satisfying the tangency conditions and the basic equation \eqref{definition_X_restricted}. The final submanifold given by the algorithm is
\begin{equation}
\mathcal{M}_3=\{(\mathbf{q}_i,q_{\{i,j\}},\mathbf{p}_i,p_{\{i,j\}})\in\Gamma\,:\,\mathcal{C}^{(0)}_{\{i,j\}}
                          =\mathcal{C}^{(1)}_{\{i,j\}}=\mathcal{C}^{(2)}_{\{i,j\}}=\mathcal{C}^{(3)}_{\{i,j\}}=0,\,\{i,j\}\in \mathcal{E}\}
\end{equation}
and the $\mathcal{M}_3$-tangent vector field is given by \eqref{Xq}, \eqref{Xp}, \eqref{Xij}.

We check now that we get the dynamics described by the original equations of motion \eqref{qequations} and \eqref{lagequations}. Indeed the equations for the integral curves of the Hamiltonian vector field $\boldsymbol{X}$ are
\begin{eqnarray}
&&\dot{\mathbf{q}}_i=\mathbf{p}_i/m\,,\hspace*{3.8cm} i\in \mathcal{V}\,,\label{equationdotq}\\
&&\dot{\mathbf{p}}_i=-\sum_{j\sim i}q_{\{i,j\}}(\mathbf{q}_i-\mathbf{q}_j)\,,\quad\hspace*{.8cm} i\in \mathcal{V}\,,\label{equationdotp}\\
&&\dot{q}_{\{1,2\}}=\dot{q}_{\{1,3\}}=\dot{q}_{\{1,4\}}=\Xi\,,\label{equationdotq12}\\
&&\dot{q}_{\{2,3\}}=\dot{q}_{\{3,4\}}=\dot{q}_{\{2,4\}}=-\Xi/3\,,\label{equationdotq24}
\end{eqnarray}
with initial conditions satisfying the constraints
\begin{equation}
\mathcal{C}^{(1)}_{\{i,j\}}=\mathcal{C}^{(2)}_{\{i,j\}}=\mathcal{C}^{(3)}_{\{i,j\}}=0\,,\quad \{i,j\}\in \mathcal{E}\,.
\label{constraints}
\end{equation}
As we can see equations \eqref{equationdotq} and \eqref{equationdotp} imply \eqref{qequations}. The constraint $\mathcal{C}^{(1)}_{\{i,j\}}=0$ is equivalent to \eqref{lagequations} and the remaining constraints are necessary for the consistency of the dynamics (in particular to choose good initial data). As we can take $\Xi$ to be an arbitrary function on $\mathcal{M}_3$ and the initial data for the $q_{\{i,j\}}$ must satisfy \eqref{qijlambda}, the solutions to \eqref{equationdotq12} and \eqref{equationdotq24} must have precisely the form given by Eqs. \eqref{q12}, \eqref{q23}.

The presence of the arbitrary function $\Xi$ in the Hamiltonian vector field $\boldsymbol{X}$ is directly related to the gauge symmetry of our system. Suppose that we pick a point $P_0$ on the submanifold $\mathcal{M}_3$, make several different choices of $\Xi$, compute the integral curves of the resulting $\boldsymbol{X}$ starting from $P_0$ at $t_0$ and take the points of these curves corresponding to the same later value of the time parameter $t>t_0$. From a physical point of view these configurations should be considered as equivalent (they certainly are, both regarding the positions of the particles and the forces acting on them at each instant of time). This leads us to the definition of \textit{gauge orbits} as constituted by all the points in phase space reachable from allowed initial data after a certain fixed time by making any possible choice of the arbitrary part of the Hamiltonian vector field defining the dynamics\footnote{Gauge orbits can be characterized also geometrically by considering the degenerate directions of the ``pulled back'' symplectic form \cite{Gotay}.}. In order to avoid the redundant description of equivalent physical configurations two options are available: \textit{gauge fixing} and the introduction of the \textit{reduced phase space}. We briefly describe them in turn.

A popular way to select the arbitrary components of the Hamiltonian vector field $\boldsymbol{X}$ (encoded in $\Xi$) is through \textit{gauge fixing}. In our example this amounts to selecting the value of the force exerted by one of the rods. This can be realized physically by substituting one of them by a spring of fixed rest length (or even removing one rod). By introducing, for instance, the additional \textit{gauge fixing} condition $\mathcal{G}:=q_{\{1,2\}}=0$ we build a submanifold $\mathcal{M}_G$ of $\mathcal{M}_3$ and fix $\Xi$ by demanding $\boldsymbol{\nabla}_{\boldsymbol{X}} \mathcal{G}=0$. This immediately gives $\Xi=0$ and removes the arbitrariness in the evolution.

The \textit{reduced phase space} is the abstract space of gauge orbits endowed with the appropriate geometric structures (in particular a symplectic form and an appropriate restriction of the Hamiltonian vector fields defining the dynamics \cite{Gotay}). In the present case --but not in generic gauge theories such as electromagnetism, Yang-Mills or general relativity-- it can be obtained by relying on the original idea by Lagrange to avoid constraint forces by working with appropriate ``generalized coordinates'' and writing the Lagrangian in terms of the kinetic and potential energy. Here the appropriate coordinates are the position of the center of mass $(x,y)\in \mathbb{R}^2$ and an angle $\theta$, i.e. a point on the unit circle $\mathbb{S}^1$. The Lagrangian is
\begin{equation}
L_R(x,y,\theta,v_x,v_y,v_\theta)=2m(v_x^2+v_y^2)+\frac{3}{2}mv_{\theta}^2\,.
\label{Lag}
\end{equation}
The system in this form is not singular, the Hamiltonian is
\begin{equation}
H_R(x,y,\theta,p_x,p_y,p_\theta)=\frac{1}{8m}(p_x^2+p_y^2)+\frac{1}{6m}p_{\theta}^2
\label{Ham}
\end{equation}
giving the unique Hamiltonian vector field
\begin{equation}
\boldsymbol{X}_{\!R}:=(X_x,X_y,X_{\theta},X_{p_x},X_{p_y},X_{p_\theta})=\Big(\frac{p_x}{4m},\frac{p_y}{4m},\frac{p_\theta}{3m},0,0,0\Big)\,.
\label{XH}
\end{equation}
It is important to mention that the reduced phase space, whose points are of the form $(x,y,\theta,p_x,p_y,p_\theta)$, is non-trivial as a manifold. Indeed it has the form $\Gamma_R=(\mathbb{R}^2\times \mathbb{S}^1)\times \mathbb{R}^3$ and, hence, is not isomorphic to a Euclidean space.

\section{Comments}\label{conclusions-sect}

As we have shown it is possible to define simple mechanical models that behave as gauge systems \textit{in a non trivial way} (some trivial examples can be found, for instance, in \cite{HT}). The main ideas are to implement the indeterminacy characteristic of hyperstatic structures in a dynamical setting and introduce constraint forces as explicit dynamical variables. The equations of motion for simple models consisting of point particles connected by \textit{ideal rods} mimic the most important features of gauge theories. In this sense they provide a useful finite dimensional analogue of gauge field theories and help as good pedagogical models to discuss other important issues such as quantization (both in the Dirac approach and by using path integral methods). It is important to mention, nonetheless, that a concrete implementation of these mechanical systems would not be subject to any indeterminacy in the individual forces exerted by the rods as a consequence of their elastic properties. In this sense the gauge behavior that we have discussed is a feature of the equations of motion in the simplified setting where the elastic properties of the rods are neglected. As far as we can see there is no analogue of this phenomenon in the standard gauge field theories.

We have discussed the obtention of the Hamiltonian description of the model by using a geometric approach. An interesting exercise is to derive the same results by following the standard method proposed by Dirac and based on the use of \textit{Poisson brackets}. In our opinion the GNH approach that we have followed is both conceptually clean and easier to use. Had we not stopped to discuss the meaning of the conditions that we have been obtaining, the computation of the Hamiltonian vector field and the constraints for our model could have been written in one page. The description of the submanifold in the full phase space where the dynamics takes place is very economical: it is seen as an algebraic manifold defined by the vanishing of simple polynomials in the canonical variables. This means, in particular, that it is defined globally in a coordinate independent way. The key concept in the obtention of the relevant manifold where the dynamics is defined and the Hamiltonian vector field is \textit{tangency}. This is both conceptually simple and easy to implement in practice.

An interesting problem for the reader would be to consider the same system after removing one rod. In this case no gauge invariance remains despite the fact that the Lagrangian is still singular. This example can help in further clarifying the relationship between singular Lagrangians and the presence of gauge symmetries.

Finally, though it may sound trivial in a sense, we would like to point out that we have adapted our notation to the graph naturally associated with our system. This approach may be useful to study the dynamics of more complicated models because the basic form of the constraints that we have obtained should generalize readily. Some of them may even be interesting as they would provide a novel way to study the rigidity of frames (a field where there are still open problems) and the dynamics of complex structures such as flexible polyhedra.

\begin{acknowledgments}

The authors want to thank Juan Margalef and Mariano Santander for their useful comments. This work has been supported by the Spanish MINECO research grants FIS2012-34379, FIS2014-57387-C3-3-P and the  Consolider-Ingenio 2010 Program CPAN (CSD2007-00042).

\end{acknowledgments}

\end{document}